\RequirePackage{lineno}
\documentclass[twocolumn,showpacs,superscriptaddress,amsmath,amssymb]{revtex4-1}
\usepackage[colorlinks,linkcolor=blue,anchorcolor=blue,citecolor=blue,urlcolor=blue]{hyperref}
\usepackage{graphicx}
\usepackage{epsfig}
\usepackage{dcolumn}
\usepackage{bm}
\usepackage{overpic}
\usepackage{color}
\usepackage{multirow}
\usepackage{subfigure}
\usepackage{lineno}

\let\oldequation\equation
\let\oldendequation\endequation

\renewenvironment{equation}
{\linenomathNonumbers\oldequation}
{\oldendequation\endlinenomath}

\begin{document}

\newcommand{\eelplm}{e^{+}e^{-} \rightarrow \Lambda_{c}^{+} \bar{\Lambda}_{c}^{-}}
\newcommand{\lplm}{\Lambda_{c}^{+} \bar{\Lambda}_{c}^{-}}
\newcommand{\eeppbar}{e^{+}e^{-} \rightarrow p\bar{p}}
\newcommand{\ee}{e^{+}e^{-}}
\newcommand{\pipi}{\pi^{+}\pi^{-}}
\newcommand{\sqs}{\sqrt{s}}
\newcommand{\ppbar}{p\bar{p}}
\newcommand{\eepipipsip}{e^{+}e^{-} \rightarrow \pi^{+}\pi^{-}\psi(2S)}
\newcommand{\pipipsip}{\pi^{+}\pi^{-}\psi(2S)}
\newcommand{\eellbar}{e^{+}e^{-} \rightarrow \Lambda\bar{\Lambda}}
\newcommand{\lam}{\Lambda_{c}^{+}}
\newcommand{\lampm}{\Lambda_{c}^{\pm}}
\newcommand{\ks}{K_{S}^{0}}
\newcommand{\Lam}{\Lambda_{c}}
\newcommand{\lambar}{\bar{\Lambda}_{c}^{-}}
\newcommand{\Lambdac}{\Lambda_{c}}
\newcommand{\lamdecay}{\Lambda_{c}^{+}\rightarrow p^{+}K^{-}\pi^{+}}
\newcommand{\lambardecay}{\bar{\Lambda}_{c}^{-}\rightarrow \bar{p}K^{+}\pi^{-}}
\newcommand{\lamdecaypks}{\Lambda_{c}^{+}\rightarrow p^{+}K_{S}^{0}}
\newcommand{\ksdecay}{K_{S}^{0}\rightarrow \pi^{+}\pi^{-}}
\newcommand{\lamdecaylampi}{\Lambda_{c}^{+}\rightarrow \Lambda\pi^{+}}
\newcommand{\lambardecaylampi}{\Lambda_{c}^{-}\rightarrow \bar{\Lambda}\pi^{-}}
\newcommand{\lambdadecay}{\Lambda\rightarrow p^{+}\pi^{-}}
\newcommand{\lamdecaypkpipi}{\Lambda_{c}^{+}\rightarrow p^{+}K^{-}\pi^{+}\pi^{0}}
\newcommand{\lamdecaypkspi}{\Lambda_{c}^{+}\rightarrow p^{+}K_{S}^{0}\pi^{0}}
\newcommand{\pidecay}{\pi^{0}\rightarrow \gamma\gamma}
\newcommand{\lamdecaylampipi}{\Lambda_{c}^{+}\rightarrow \Lambda\pi^{+}\pi^{0}}
\newcommand{\lamdecaypkspipi}{\Lambda_{c}^{+}\rightarrow p^{+}K_{S}^{0}\pi^{+}\pi^{-}}
\newcommand{\lamdecaylampipipi}{\Lambda_{c}^{+}\rightarrow \Lambda\pi^{+}\pi^{+}\pi^{-}}
\newcommand{\lamdecaysigmapi}{\Lambda_{c}^{+}\rightarrow \Sigma^{0}\pi^{+}}
\newcommand{\sigmadecay}{\Sigma^{0}\rightarrow \Lambda\gamma}
\newcommand{\lamdecaysigmapipi}{\Lambda_{c}^{+}\rightarrow \Sigma^{+}\pi^{+}\pi^{-}}
\newcommand{\sigmapdecay}{\Sigma^{+}\rightarrow p\pi^{0}}
\newcommand{\mbc}{M_{\textmd{BC}}}
\newcommand{\mbcp}{M_{\textmd{BC}}^{+}}
\newcommand{\mbcm}{M_{\textmd{BC}}^{-}}
\newcommand{\mbcpm}{M_{\textmd{BC}}^{\pm}}
\newcommand{\mbcmp}{M_{\textmd{BC}}^{\mp}}
\newcommand{\mbcmax}{M_{\textmd{BC}}^{\textmd{max}}}
\newcommand{\dele}{\Delta E}
\newcommand{\delep}{\Delta E_{+}}
\newcommand{\delem}{\Delta E_{-}}
\newcommand{\delepm}{\Delta E_{\pm}}
\newcommand{\delemp}{\Delta E_{\mp}}
\newcommand{\delemin}{\Delta E_{\textmd{min}}}
\newcommand{\ebm}{E_{\textmd{beam}}}
\newcommand{\gev}{\mathrm{GeV}}
\newcommand{\mev}{\mathrm{MeV}}
\newcommand{\gevc}{\mathrm{GeV}/c}
\newcommand{\mevc}{\mathrm{MeV}/c}
\newcommand{\gevcc}{\mathrm{GeV}/c^2}
\newcommand{\mevcc}{\mathrm{MeV}/c^2}
\newcommand{\modeB}{ \bar{p}K^{+}\pi^{-}}
\newcommand{\modeI}{ pK^{-}\pi^{+}}
\newcommand{\modeII}{ pK_{S}^{0}}
\newcommand{\modeIII}{ \Lambda\pi^{+}}
\newcommand{\modeIV}{ pK^{-}\pi^{+}\pi^{0}}
\newcommand{\modeV}{ pK_{S}^{0}\pi^{0}}
\newcommand{\modeVI}{ \Lambda\pi^{+}\pi^{0}}
\newcommand{\modeVII}{ pK_{S}^{0}\pi^{+}\pi^{-}}
\newcommand{\modeVIII}{ \Lambda\pi^{+}\pi^{+}\pi^{-}}
\newcommand{\modeIX}{ \Sigma^{0}\pi^{+}}
\newcommand{\modeX}{ \Sigma^{+}\pi^{+}\pi^{-}}
\newcommand{\lambdac}{\Lambda_{c}}
\newcommand{\amodeI}{ \bar{p}K^{+}\pi^{-}}
\newcommand{\amodeII}{ \bar{p}K_{S}^{0}}
\newcommand{\amodeIII}{ \bar{\Lambda}\pi^{-}}
\newcommand{\amodeIV}{ \bar{p}K^{+}\pi^{-}\pi^{0}}
\newcommand{\amodeV}{ \bar{p}K_{S}^{0}\pi^{0}}
\newcommand{\amodeVI}{ \bar{\Lambda}\pi^{-}\pi^{0}}
\newcommand{\amodeVII}{ \bar{p}K_{S}^{0}\pi^{-}\pi^{+}}
\newcommand{\amodeVIII}{ \bar{\Lambda}\pi^{-}\pi^{+}\pi^{+}}
\newcommand{\amodeIX}{ \bar{\Sigma^{0}}\pi^{-}}
\newcommand{\amodeX}{ \bar{\Sigma}^{-}\pi^{-}\pi^{+}}

\newcommand{\ndatlamp}{ N^{\textmd{data}}_{\lam} }
\newcommand{\ndatlamm}{ N^{\textmd{data}}_{\lambar}  }
\newcommand{\ndatmodei}{ N^{\textmd{data}}_{i}  }
\newcommand{\ndat}{ N^{\textmd{data}}  }
\newcommand{\nposcrt}{ N_{\textmd{crt}}^{+}  }
\newcommand{\nnegcrt}{ N_{\textmd{crt}}^{-}  }
\newcommand{\navecrt}{ \bar{N}^{\textmd{crt}}  }
\newcommand{\efflamp}{ \varepsilon_{\lam}  }
\newcommand{\efflamm}{  \varepsilon_{\lambar}  }
\newcommand{\effmodei}{  \varepsilon_{i}  }
\newcommand{\eff}{  \varepsilon  }
\newcommand{\csbornlamp}{ \sigma_{\lam}^{\textmd{Born}} }
\newcommand{\csbornlamm}{ \sigma_{\lambar}^{\textmd{Born}} }
\newcommand{\csborn}{ \sigma_{\textmd{Born}} }
\newcommand{\csobslamp}{ \sigma_{\lam}^{\textmd{obs}} }
\newcommand{\csobslamm}{ \sigma_{\lambar}^{\textmd{obs}} }
\newcommand{\csobs}{ \sigma_{\textmd{obs}} }
\newcommand{\cspm}{ \sigma_{\pm} }
\newcommand{\csp}{ \sigma_{+} }
\newcommand{\csm}{ \sigma_{-} }
\newcommand{\brf}{ \mathcal{B}  }
\newcommand{\brfpm}{ \mathcal{B}_{\pm}  }
\newcommand{\brfp}{ \mathcal{B}_{+}  }
\newcommand{\brfm}{ \mathcal{B}_{-}  }
\newcommand{\brfmodei}{ \mathcal{B}_{i}  }
\newcommand{\lint}{ \mathcal{L}_{\textmd{int}}  }

\newcommand{\nstcosk}{ N^{k}_{\textmd{ST}} }
\newcommand{\nstcospk}{ N^{+,k}_{\textmd{ST}} }
\newcommand{\nstcosmk}{ N^{-,k}_{\textmd{ST}} }
\newcommand{\nstcospmk}{ N^{\pm,k}_{\textmd{ST}} }
\newcommand{\ncrtbink}{ N_{\textmd{crt}}^{k}  }
\newcommand{\ncrtcospk}{ N_{\textmd{crt}}^{+,k}  }
\newcommand{\ncrtcosmk}{ N_{\textmd{crt}}^{-,k}  }
\newcommand{\ncrtcospmk}{ N_{\textmd{crt}}^{\pm,k}  }

\newcommand{\ncrtcosavek}{ \bar{N}_{\textmd{crt}}^{k}  }
\newcommand{\ndatbink}{ N^{\textmd{data}}_{k}  }
\newcommand{\effstcosk}{ \varepsilon^{k}_{\textmd{ST}} }
\newcommand{\effstcospk}{ \varepsilon^{+,k}_{\textmd{ST}} }
\newcommand{\effstcosmk}{  \varepsilon^{-,k}_{\textmd{ST}} }
\newcommand{\effstcospmk}{ \varepsilon^{\pm,k}_{\textmd{ST}} }
\newcommand{\nbin}{ N_{\textmd{bin}} }
\newcommand{\linteff}{ \mathcal{L}_{\textmd{int}}^{\textmd{eff}} }
\newcommand{\linteqv}{ \mathcal{L}_{\textmd{int}}^{\textmd{eqv}} }

\newcommand{\ncrtcospkbar}{ N_{\textmd{crt}}^{+,\nbin-k}  }
\newcommand{\ncrtcosmkbar}{ N_{\textmd{crt}}^{-,\nbin-k}  }
\newcommand{\ncrtcospmkbar}{ N_{\textmd{crt}}^{\pm,\nbin-k}  }
\newcommand{\fbAsypmk}{ \mathcal{A}_{FB}^{\pm,k} }
\newcommand{\fbAsypk}{ \mathcal{A}_{FB}^{+,k} }
\newcommand{\fbAsymk}{ \mathcal{A}_{FB}^{-,k} }
\newcommand{\fbAsycrtk}{ \mathcal{A}_{FB}^{\textmd{crt,k}} }

\newcommand{\ndatmodeik}{ N^{\textmd{data}}_{i,k}  }
\newcommand{\effavelampk}{ \bar{\varepsilon}_{\lam,k}  }
\newcommand{\effavelammk}{ \bar{\varepsilon}_{\lambar,k}  }
\newcommand{\effavek}{ \bar{\varepsilon}_{k}  }
\newcommand{\effmodeik}{  \varepsilon_{i,k}  }
\newcommand{\effbink}{ \varepsilon_{k}  }
\newcommand{\nmode}{ N_{\textmd{mode}} }

\newcommand{\ngen}{ N_{\textmd{gen}} }
\newcommand{\nsur}{ N_{\textmd{sur}} }
\newcommand{\nlplm}{N_{\lplm}}
\newcommand{\nst}{N_{\textmd{ST}}}
\newcommand{\ndt}{N_{\textmd{DT}}}
\newcommand{\effst}{\varepsilon_{\textmd{ST}}}
\newcommand{\effdt}{\varepsilon_{\textmd{DT}}}

\newcommand{\nstpm}{N_{\textmd{ST}}^{\pm}}
\newcommand{\effstpm}{\varepsilon_{\textmd{ST}}^{\pm}}

\newcommand{\nstp}{N_{\textmd{ST}}^{+}}
\newcommand{\effstp}{\varepsilon_{\textmd{ST}}^{+}}
\newcommand{\nstm}{N_{\textmd{ST}}^{-}}
\newcommand{\effstm}{\varepsilon_{\textmd{ST}}^{-}}

\newcommand{\qqbar}{ q\bar{q} }
\newcommand{\alfs}{\alpha_{s}}
\newcommand{\alfmz}{\alpha(M_{Z}^{2})}
\newcommand{\tata}{\tau^{+}\tau^{-}}
\newcommand{\twopho}{\ee+X}
\newcommand{\lmlp}{\lam\lambar}
\newcommand{\fvp}{f_{\textmd{VP}}}
\newcommand{\fisr}{f_{\textmd{ISR}}}
\newcommand{\sigmodei}{\sigma_{i}}
\newcommand{\crosecerr}{\Delta\sigma}
\newcommand{\avesigerr}{\Delta\sigma}
\newcommand{\avesigstaerr}{\Delta\sigma_{\textmd{stat}}}
\newcommand{\avesigsyserr}{\Delta\sigma_{\textmd{syst}}}
\newcommand{\errmat}{\mathbf{M}^{\sigma}}
\newcommand{\errmatsta}{\mathbf{M}^{\sigma}_{\textmd{stat}}}
\newcommand{\errmatsys}{\mathbf{M}^{\sigma}_{\textmd{syst}}}
\newcommand{\reladiff}{\Delta_{\textmd{rel}}}

\newcommand{\sigs}{\sigma(s)}
\newcommand{\costh}{\cos\theta}
\newcommand{\cossqth}{\cos^{2}\theta}
\newcommand{\gamvir}{\gamma^{\ast}}

\newcommand{\gE}{ G_{E} }
\newcommand{\gM}{ G_{M} }
\newcommand{\GE}{ |G_{E}| }
\newcommand{\GM}{ |G_{M}| }
\newcommand{\Geff}{ |G_{\textmd{eff}}| }
\newcommand{\gmmod}{ |G_{M}| }
\newcommand{\gEgM}{ |G_{E}/G_{M}| }
\newcommand{\ratios}{ |G_{E}/G_{M}| }
\newcommand{\alplam}{ \alpha_{\Lam} }
\newcommand{\alplampm}{ \alpha_{\Lam}^{\pm} }
\newcommand{\alplamp}{ \alpha_{\lam} }
\newcommand{\alplamm}{ \alpha_{\lambar} }
\newcommand{\alpisr}{ \alpha_{\textmd{ISR}} }
\newcommand{\alpisrpm}{ \alpha_{\textmd{ISR}}^{\pm} }
\newcommand{\alpisrp}{ \alpha_{\textmd{ISR}}^{+} }
\newcommand{\alpisrm}{ \alpha_{\textmd{ISR}}^{-} }
\newcommand{\Fisr}{F_{\textmd{ISR}}}
\newcommand{\Fisrpm}{F_{\textmd{ISR}}^{\pm}}
\newcommand{\Fisrpmk}{F_{\textmd{ISR}}^{\pm,k}}
\newcommand{\Fisrpk}{F_{\textmd{ISR}}^{+,k}}
\newcommand{\Fisrmk}{F_{\textmd{ISR}}^{-,k}}
\newcommand{\Fisrppmk}{F_{\textmd{ISR}}^{\prime\pm,k}}

\newcommand{\fisrM}{ 0.445 }
\newcommand{\LintM}{ 48.93 }
\newcommand{\lintM}{ 48.9 }
\newcommand{\PkpiBornCSPM}{ 244\pm18\pm46 }
\newcommand{\PkpiBornCSMM}{ 237\pm18\pm45 }
\newcommand{\PkpiObsvCSPM}{ 115\pm8\pm22 }
\newcommand{\PkpiObsvCSMM}{ 112\pm8\pm21 }

\newcommand{\fisrN}{ 0.661 }
\newcommand{\LintN}{ 8.84 }
\newcommand{\lintN}{ 8.84 }
\newcommand{\PkpiBornCSPN}{ 162\pm30\pm6 }
\newcommand{\PkpiBornCSMN}{ 231\pm35\pm8 }
\newcommand{\PkpiObsvCSPN}{ 113\pm21\pm4 }
\newcommand{\PkpiObsvCSMN}{ 161\pm24\pm6 }

\newcommand{\fisrO}{ 0.714 }
\newcommand{\LintO}{ 8.49 }
\newcommand{\lintO}{ 8.49 }
\newcommand{\PkpiBornCSPO}{ 237\pm36\pm9 }
\newcommand{\PkpiBornCSMO}{ 237\pm35\pm9 }
\newcommand{\PkpiObsvCSPO}{ 178\pm27\pm7 }
\newcommand{\PkpiObsvCSMO}{ 178\pm27\pm7 }

\newcommand{\fisrP}{ 0.7428 }
\newcommand{\LintP}{ 586.89 }
\newcommand{\lintP}{ 586.9 }
\newcommand{\PkpiBornCSPP}{ 212.3\pm4.2\pm7.3 }
\newcommand{\PkpiBornCSMP}{ 219.4\pm4.0\pm7.5 }
\newcommand{\PkpiObsvCSPP}{ 166.3\pm3.3\pm5.7 }
\newcommand{\PkpiObsvCSMP}{ 171.9\pm3.1\pm5.9 }

\newcommand{\fisrA}{ 0.7657 }
\newcommand{\LintA}{ 103.65 }
\newcommand{\lintA}{ 103.7 }
\newcommand{\PkpiBornCSPA}{ 214.6\pm10.1\pm7.3 }
\newcommand{\PkpiBornCSMA}{ 202.8\pm9.5\pm6.9 }
\newcommand{\PkpiObsvCSPA}{ 173.2\pm8.1\pm5.9 }
\newcommand{\PkpiObsvCSMA}{ 163.8\pm7.7\pm5.6 }

\newcommand{\fisrB}{ 0.7868 }
\newcommand{\LintB}{ 521.53 }
\newcommand{\lintB}{ 521.5 }
\newcommand{\PkpiBornCSPB}{ 207.0\pm4.4\pm7.0 }
\newcommand{\PkpiBornCSMB}{ 205.7\pm4.3\pm7.0 }
\newcommand{\PkpiObsvCSPB}{ 171.7\pm3.7\pm5.8 }
\newcommand{\PkpiObsvCSMB}{ 170.7\pm3.6\pm5.8 }

\newcommand{\fisrC}{ 0.8001 }
\newcommand{\LintC}{ 551.65 }
\newcommand{\lintC}{ 551.6 }
\newcommand{\PkpiBornCSPC}{ 208.1\pm4.3\pm7.1 }
\newcommand{\PkpiBornCSMC}{ 202.3\pm4.1\pm6.9 }
\newcommand{\PkpiObsvCSPC}{ 175.5\pm3.6\pm6.0 }
\newcommand{\PkpiObsvCSMC}{ 170.7\pm3.5\pm5.8 }

\newcommand{\fisrD}{ 0.8193 }
\newcommand{\LintD}{ 529.43 }
\newcommand{\lintD}{ 529.4 }
\newcommand{\PkpiBornCSPD}{ 206.6\pm4.3\pm7.1 }
\newcommand{\PkpiBornCSMD}{ 194.8\pm4.0\pm6.7 }
\newcommand{\PkpiObsvCSPD}{ 178.4\pm3.7\pm6.1 }
\newcommand{\PkpiObsvCSMD}{ 168.3\pm3.5\pm5.7 }

\newcommand{\fisrE}{ 0.8422 }
\newcommand{\LintE}{ 1667.39 }
\newcommand{\lintE}{ 1667.4 }
\newcommand{\PkpiBornCSPE}{ 186.1\pm2.3\pm6.3 }
\newcommand{\PkpiBornCSME}{ 190.2\pm2.2\pm6.5 }
\newcommand{\PkpiObsvCSPE}{ 165.3\pm2.0\pm5.6 }
\newcommand{\PkpiObsvCSME}{ 168.9\pm2.0\pm5.8 }

\newcommand{\fisrF}{ 0.8668 }
\newcommand{\LintF}{ 535.54 }
\newcommand{\lintF}{ 535.5 }
\newcommand{\PkpiBornCSPF}{ 169.1\pm3.8\pm5.9 }
\newcommand{\PkpiBornCSMF}{ 175.6\pm3.7\pm6.2 }
\newcommand{\PkpiObsvCSPF}{ 154.6\pm3.5\pm5.4 }
\newcommand{\PkpiObsvCSMF}{ 160.5\pm3.4\pm5.6 }

\newcommand{\fisrG}{ 0.9300 }
\newcommand{\LintG}{ 163.87 }
\newcommand{\lintG}{ 163.9 }
\newcommand{\PkpiBornCSPG}{ 121.0\pm5.9\pm5.0 }
\newcommand{\PkpiBornCSMG}{ 125.9\pm5.8\pm5.2 }
\newcommand{\PkpiObsvCSPG}{ 118.7\pm5.8\pm4.9 }
\newcommand{\PkpiObsvCSMG}{ 123.5\pm5.7\pm5.1 }

\newcommand{\fisrH}{ 0.9072 }
\newcommand{\LintH}{ 366.55 }
\newcommand{\lintH}{ 366.6 }
\newcommand{\PkpiBornCSPH}{ 134.2\pm4.2\pm4.6 }
\newcommand{\PkpiBornCSMH}{ 123.7\pm3.9\pm4.3 }
\newcommand{\PkpiObsvCSPH}{ 128.4\pm4.0\pm4.4 }
\newcommand{\PkpiObsvCSMH}{ 118.4\pm3.7\pm4.1 }

\newcommand{\fisrI}{ 0.9185 }
\newcommand{\LintI}{ 511.47 }
\newcommand{\lintI}{ 511.5 }
\newcommand{\PkpiBornCSPI}{ 124.2\pm3.4\pm4.3 }
\newcommand{\PkpiBornCSMI}{ 123.7\pm3.3\pm4.3 }
\newcommand{\PkpiObsvCSPI}{ 120.4\pm3.3\pm4.2 }
\newcommand{\PkpiObsvCSMI}{ 119.9\pm3.2\pm4.1 }

\newcommand{\fisrJ}{ 0.9662 }
\newcommand{\LintJ}{ 525.16 }
\newcommand{\lintJ}{ 525.2 }
\newcommand{\PkpiBornCSPJ}{ 84.9\pm2.9\pm2.9 }
\newcommand{\PkpiBornCSMJ}{ 84.6\pm2.7\pm2.9 }
\newcommand{\PkpiObsvCSPJ}{ 86.6\pm3.0\pm3.0 }
\newcommand{\PkpiObsvCSMJ}{ 86.3\pm2.8\pm3.0 }

\newcommand{\fisrK}{ 0.9583 }
\newcommand{\LintK}{ 207.82 }
\newcommand{\lintK}{ 207.8 }
\newcommand{\PkpiBornCSPK}{ 96.9\pm4.7\pm3.5 }
\newcommand{\PkpiBornCSMK}{ 99.3\pm4.6\pm3.5 }
\newcommand{\PkpiObsvCSPK}{ 98.1\pm4.8\pm3.5 }
\newcommand{\PkpiObsvCSMK}{ 100.6\pm4.7\pm3.6 }

\newcommand{\fisrL}{ 0.9557 }
\newcommand{\LintL}{ 159.28 }
\newcommand{\lintL}{ 159.3 }
\newcommand{\PkpiBornCSPL}{ 87.1\pm5.1\pm3.1 }
\newcommand{\PkpiBornCSML}{ 92.2\pm5.1\pm3.3 }
\newcommand{\PkpiObsvCSPL}{ 88.0\pm5.2\pm3.1 }
\newcommand{\PkpiObsvCSML}{ 93.1\pm5.2\pm3.3 }

\newcommand{\PkpiBornCSCM}{ 241\pm13\pm45 }
\newcommand{\PkpiObsvCSCM}{ 113\pm6\pm21 }
\newcommand{\PkpiBornGefM}{ 99\pm3\pm9 }

\newcommand{\PkpiBornCSCN}{ 191\pm23\pm7 }
\newcommand{\PkpiObsvCSCN}{ 133\pm16\pm5 }
\newcommand{\PkpiBornGefN}{ 67\pm4\pm1 }

\newcommand{\PkpiBornCSCO}{ 237\pm25\pm9 }
\newcommand{\PkpiObsvCSCO}{ 178\pm19\pm7 }
\newcommand{\PkpiBornGefO}{ 63\pm3\pm1 }

\newcommand{\PkpiBornCSCP}{ 215.8\pm2.9\pm7.3 }
\newcommand{\PkpiObsvCSCP}{ 169.1\pm2.3\pm5.7 }
\newcommand{\PkpiBornGefP}{ 54.3\pm0.4\pm0.9 }

\newcommand{\PkpiBornCSCA}{ 208.4\pm6.9\pm7.0 }
\newcommand{\PkpiObsvCSCA}{ 168.2\pm5.6\pm5.7 }
\newcommand{\PkpiBornGefA}{ 49.2\pm0.8\pm0.8 }

\newcommand{\PkpiBornCSCB}{ 206.4\pm3.1\pm6.9 }
\newcommand{\PkpiObsvCSCB}{ 171.2\pm2.6\pm5.8 }
\newcommand{\PkpiBornGefB}{ 45.5\pm0.3\pm0.8 }

\newcommand{\PkpiBornCSCC}{ 205.1\pm3.0\pm6.9 }
\newcommand{\PkpiObsvCSCC}{ 173.0\pm2.5\pm5.8 }
\newcommand{\PkpiBornGefC}{ 43.4\pm0.3\pm0.7 }

\newcommand{\PkpiBornCSCD}{ 200.3\pm2.9\pm6.8 }
\newcommand{\PkpiObsvCSCD}{ 173.1\pm2.5\pm5.8 }
\newcommand{\PkpiBornGefD}{ 40.6\pm0.3\pm0.7 }

\newcommand{\PkpiBornCSCE}{ 188.1\pm1.6\pm6.3 }
\newcommand{\PkpiObsvCSCE}{ 167.1\pm1.4\pm5.6 }
\newcommand{\PkpiBornGefE}{ 37.7\pm0.2\pm0.6 }

\newcommand{\PkpiBornCSCF}{ 172.3\pm2.7\pm6.0 }
\newcommand{\PkpiObsvCSCF}{ 157.5\pm2.5\pm5.5 }
\newcommand{\PkpiBornGefF}{ 35.1\pm0.3\pm0.6 }

\newcommand{\PkpiBornCSCG}{ 123.5\pm4.2\pm5.0 }
\newcommand{\PkpiObsvCSCG}{ 121.1\pm4.1\pm4.9 }
\newcommand{\PkpiBornGefG}{ 28.2\pm0.5\pm0.6 }

\newcommand{\PkpiBornCSCH}{ 128.5\pm2.8\pm4.4 }
\newcommand{\PkpiObsvCSCH}{ 123.0\pm2.7\pm4.2 }
\newcommand{\PkpiBornGefH}{ 28.5\pm0.3\pm0.5 }

\newcommand{\PkpiBornCSCI}{ 124.0\pm2.4\pm4.2 }
\newcommand{\PkpiObsvCSCI}{ 120.1\pm2.3\pm4.1 }
\newcommand{\PkpiBornGefI}{ 27.2\pm0.3\pm0.5 }

\newcommand{\PkpiBornCSCJ}{ 84.8\pm2.0\pm2.9 }
\newcommand{\PkpiObsvCSCJ}{ 86.5\pm2.0\pm2.9 }
\newcommand{\PkpiBornGefJ}{ 21.6\pm0.3\pm0.4 }

\newcommand{\PkpiBornCSCK}{ 98.1\pm3.3\pm3.5 }
\newcommand{\PkpiObsvCSCK}{ 99.3\pm3.3\pm3.5 }
\newcommand{\PkpiBornGefK}{ 22.4\pm0.4\pm0.4 }

\newcommand{\PkpiBornCSCL}{ 89.6\pm3.6\pm3.1 }
\newcommand{\PkpiObsvCSCL}{ 90.5\pm3.7\pm3.2 }
\newcommand{\PkpiBornGefL}{ 21.2\pm0.4\pm0.4 }

\newcommand{\brfpval}{(6.53\pm0.21)\%}
\newcommand{\brfmval}{(6.79\pm0.22)\%}
\newcommand{\brfval }{(6.66\pm0.22)\%}
\newcommand{\ndtval }{1007\pm32}
\newcommand{\brfvalall }{(6.66\pm0.22\pm0.05)\%}

\newcommand{\FinalpHaI}{ -0.13\pm0.12\pm0.08 }
\newcommand{\FinRatioI}{ 1.14\pm0.14\pm0.07 }

\newcommand{\FinalpHaO}{ -0.20\pm0.04\pm0.02 }
\newcommand{\FinRatioO}{ 1.23\pm0.05\pm0.03 }


\newcommand{\pkpiFinalpHaM}{ -0.25\pm0.14\pm0.73 }
\newcommand{\pkpiFinscaHaM}{ 19.8\pm1.5\pm9.2 }
\newcommand{\pkpiFinRatioM}{ 1.29\pm0.19\pm0.70 }
\newcommand{\pkpiFinGmmodM}{ 88.8\pm10.9\pm17.0 }
\newcommand{\pkpiFinGemodM}{ 115.0\pm4.6\pm64.4 }

\newcommand{\pkpiFinalpHaP}{ -0.23\pm0.04\pm0.01 }
\newcommand{\pkpiFinscaHaP}{ 356.8\pm6.8\pm1.8 }
\newcommand{\pkpiFinRatioP}{ 1.27\pm0.05\pm0.02 }
\newcommand{\pkpiFinGmmodP}{ 49.5\pm1.6\pm1.7 }
\newcommand{\pkpiFinGemodP}{ 62.8\pm0.7\pm2.1 }

\newcommand{\pkpiFinalpHaA}{ -0.26\pm0.09\pm0.01 }
\newcommand{\pkpiFinscaHaA}{ 61.6\pm2.8\pm1.2 }
\newcommand{\pkpiFinRatioA}{ 1.31\pm0.12\pm0.01 }
\newcommand{\pkpiFinGmmodA}{ 43.5\pm3.3\pm1.5 }
\newcommand{\pkpiFinGemodA}{ 57.1\pm1.4\pm2.0 }

\newcommand{\pkpiFinalpHaB}{ -0.21\pm0.04\pm0.01 }
\newcommand{\pkpiFinscaHaB}{ 318.9\pm6.7\pm3.0 }
\newcommand{\pkpiFinRatioB}{ 1.25\pm0.06\pm0.01 }
\newcommand{\pkpiFinGmmodB}{ 41.8\pm1.5\pm1.5 }
\newcommand{\pkpiFinGemodB}{ 52.1\pm0.7\pm1.8 }

\newcommand{\pkpiFinalpHaC}{ -0.09\pm0.05\pm0.01 }
\newcommand{\pkpiFinscaHaC}{ 328.5\pm6.9\pm2.6 }
\newcommand{\pkpiFinRatioC}{ 1.11\pm0.05\pm0.01 }
\newcommand{\pkpiFinGmmodC}{ 41.8\pm1.4\pm1.4 }
\newcommand{\pkpiFinGemodC}{ 46.5\pm0.7\pm1.6 }

\newcommand{\pkpiFinalpHaD}{ -0.02\pm0.05\pm0.01 }
\newcommand{\pkpiFinscaHaD}{ 307.8\pm6.6\pm1.5 }
\newcommand{\pkpiFinRatioD}{ 1.04\pm0.05\pm0.01 }
\newcommand{\pkpiFinGmmodD}{ 40.2\pm1.4\pm1.4 }
\newcommand{\pkpiFinGemodD}{ 41.7\pm0.7\pm1.4 }

\newcommand{\pkpiFinalpHaE}{ 0.15\pm0.03\pm0.01 }
\newcommand{\pkpiFinscaHaE}{ 881.7\pm11.4\pm6.0 }
\newcommand{\pkpiFinRatioE}{ 0.88\pm0.03\pm0.01 }
\newcommand{\pkpiFinGmmodE}{ 39.2\pm0.8\pm1.3 }
\newcommand{\pkpiFinGemodE}{ 34.4\pm0.5\pm1.2 }

\newcommand{\pkpiFinalpHaF}{ 0.34\pm0.07\pm0.01 }
\newcommand{\pkpiFinscaHaF}{ 252.8\pm6.2\pm2.1 }
\newcommand{\pkpiFinRatioF}{ 0.72\pm0.06\pm0.01 }
\newcommand{\pkpiFinGmmodF}{ 38.2\pm1.4\pm1.3 }
\newcommand{\pkpiFinGemodF}{ 27.6\pm1.2\pm1.0 }

\newcommand{\pkpiFinalpHaG}{ 0.49\pm0.16\pm0.03 }
\newcommand{\pkpiFinscaHaG}{ 56.6\pm3.1\pm1.3 }
\newcommand{\pkpiFinRatioG}{ 0.61\pm0.13\pm0.02 }
\newcommand{\pkpiFinGmmodG}{ 31.4\pm2.4\pm1.3 }
\newcommand{\pkpiFinGemodG}{ 19.2\pm2.6\pm1.0 }

\newcommand{\pkpiFinalpHaH}{ 0.42\pm0.10\pm0.01 }
\newcommand{\pkpiFinscaHaH}{ 131.6\pm4.7\pm1.7 }
\newcommand{\pkpiFinRatioH}{ 0.66\pm0.08\pm0.01 }
\newcommand{\pkpiFinGmmodH}{ 31.4\pm1.6\pm1.1 }
\newcommand{\pkpiFinGemodH}{ 20.8\pm1.5\pm0.8 }

\newcommand{\pkpiFinalpHaI}{ 0.17\pm0.07\pm0.01 }
\newcommand{\pkpiFinscaHaI}{ 193.4\pm5.6\pm2.7 }
\newcommand{\pkpiFinRatioI}{ 0.88\pm0.07\pm0.01 }
\newcommand{\pkpiFinGmmodI}{ 28.2\pm1.2\pm1.0 }
\newcommand{\pkpiFinGemodI}{ 24.8\pm0.9\pm0.9 }

\newcommand{\pkpiFinalpHaJ}{ 0.38\pm0.10\pm0.01 }
\newcommand{\pkpiFinscaHaJ}{ 134.4\pm4.9\pm2.1 }
\newcommand{\pkpiFinRatioJ}{ 0.71\pm0.09\pm0.01 }
\newcommand{\pkpiFinGmmodJ}{ 23.4\pm1.3\pm0.8 }
\newcommand{\pkpiFinGemodJ}{ 16.7\pm1.2\pm0.6 }

\newcommand{\pkpiFinalpHaK}{ 0.62\pm0.17\pm0.01 }
\newcommand{\pkpiFinscaHaK}{ 56.5\pm3.1\pm1.1 }
\newcommand{\pkpiFinRatioK}{ 0.52\pm0.15\pm0.01 }
\newcommand{\pkpiFinGmmodK}{ 25.3\pm1.9\pm0.9 }
\newcommand{\pkpiFinGemodK}{ 13.2\pm2.7\pm0.5 }

\newcommand{\pkpiFinalpHaL}{ 0.63\pm0.21\pm0.01 }
\newcommand{\pkpiFinscaHaL}{ 40.0\pm2.6\pm0.8 }
\newcommand{\pkpiFinRatioL}{ 0.52\pm0.18\pm0.01 }
\newcommand{\pkpiFinGmmodL}{ 24.1\pm2.2\pm0.9 }
\newcommand{\pkpiFinGemodL}{ 12.5\pm3.1\pm0.5 }


\newcommand{\ENERGYMT}{ 4.5745 }
\newcommand{\ENERGYNT}{ 4.5800 }
\newcommand{\ENERGYOT}{ 4.5900 }
\newcommand{\ENERGYPT}{ 4.5995 }
\newcommand{\ENERGYAT}{ 4.6119 }
\newcommand{\ENERGYBT}{ 4.6280 }
\newcommand{\ENERGYCT}{ 4.6409 }
\newcommand{\ENERGYDT}{ 4.6612 }
\newcommand{\ENERGYET}{ 4.6819 }
\newcommand{\ENERGYFT}{ 4.6988 }
\newcommand{\ENERGYGT}{ 4.7397 }
\newcommand{\ENERGYHT}{ 4.7500 }
\newcommand{\ENERGYIT}{ 4.7805 }
\newcommand{\ENERGYJT}{ 4.8431 }
\newcommand{\ENERGYKT}{ 4.9180 }
\newcommand{\ENERGYLT}{ 4.9509 }

\newcommand{\ENERGYM}{ 4.5745 }
\newcommand{\ENERGYN}{ 4.58 }
\newcommand{\ENERGYO}{ 4.90 }
\newcommand{\ENERGYP}{ 4.60 }
\newcommand{\ENERGYA}{ 4.61 }
\newcommand{\ENERGYB}{ 4.63 }
\newcommand{\ENERGYC}{ 4.64 }
\newcommand{\ENERGYD}{ 4.66 }
\newcommand{\ENERGYE}{ 4.68 }
\newcommand{\ENERGYF}{ 4.70 }
\newcommand{\ENERGYG}{ 4.74 }
\newcommand{\ENERGYH}{ 4.75 }
\newcommand{\ENERGYI}{ 4.78 }
\newcommand{\ENERGYJ}{ 4.84 }
\newcommand{\ENERGYK}{ 4.92 }
\newcommand{\ENERGYL}{ 4.95 }


\title{\boldmath Measurement of the Energy-Dependent Electromagnetic Form Factors \\of a Charmed Baryon}

\author{
\begin{small}
\begin{center}
M.~Ablikim$^{1}$, M.~N.~Achasov$^{5,b}$, P.~Adlarson$^{74}$, X.~C.~Ai$^{80}$, R.~Aliberti$^{35}$, A.~Amoroso$^{73A,73C}$, M.~R.~An$^{39}$, Q.~An$^{70,57}$, Y.~Bai$^{56}$, O.~Bakina$^{36}$
% ...
\\
\vspace{0.2cm}
(BESIII Collaboration)\\
\vspace{0.2cm} {\it
$^{1}$ Institute of High Energy Physics, Beijing 100049, People's Republic of China\\
% ...
$^{5}$ Budker Institute of Nuclear Physics SB RAS (BINP), Novosibirsk 630090, Russia\\
% ...
$^{35}$ Johannes Gutenberg University of Mainz, Johann-Joachim-Becher-Weg 45, D-55099 Mainz, Germany\\
$^{36}$ Joint Institute for Nuclear Research, 141980 Dubna, Moscow region, Russia\\
% ...
$^{39}$ Liaoning Normal University, Dalian 116029, People's Republic of China\\
% ...
$^{56}$ Southeast University, Nanjing 211100, People's Republic of China\\
$^{57}$ State Key Laboratory of Particle Detection and Electronics, Beijing 100049, Hefei 230026, People's Republic of China\\
% ...
$^{70}$ University of Science and Technology of China, Hefei 230026, People's Republic of China\\
% ...
$^{73}$ University of Turin and INFN, (A)University of Turin, I-10125, Turin, Italy; (B)University of Eastern Piedmont, I-15121, Alessandria, Italy; (C)INFN, I-10125, Turin, Italy\\
$^{74}$ Uppsala University, Box 516, SE-75120 Uppsala, Sweden\\
% ...
$^{80}$ Zhengzhou University, Zhengzhou 450001, People's Republic of China\\
\vspace{0.2cm}
% ...
$^{b}$ Also at the Novosibirsk State University, Novosibirsk, 630090, Russia\\
% ...
}
\end{center}
\vspace{0.4cm}
\vspace{0.4cm}
\end{small}
}

\noaffiliation{}

\date{\today}

\begin{abstract}
We study the process $\eelplm$ at twelve center-of-mass energies from $\ENERGYAT$ to $\ENERGYLT~\gev$ using data samples collected by the BESIII detector at the BEPCII collider. 
The Born cross sections and effective form factors ($\Geff$) are determined with unprecedented precision after combining the single and double-tag methods based on the decay process $\lam\to\modeI$. 
Flat cross sections around $4.63~\gev$ are obtained and no indication of the resonant structure $Y(4630)$, as reported by Belle, is found. 
In addition, no oscillatory behavior is discerned in the $\Geff$ energy-dependence of $\lam$, in contrast to what is seen for the proton and neutron cases. 
Analyzing the cross section together with the polar-angle distribution of the $\lam$ baryon at each energy point, the moduli of electric and magnetic form factors ($\GE$ and $\GM$) are extracted and separated. For the first time, the energy-dependence of the form factor ratio $\gEgM$ is observed, which can be well described by an oscillatory function. 
\end{abstract}

\maketitle

One of the most challenging aspects of the Standard Model of particle physics is to understand quantitatively how the strong interaction, as described by the fundamental theory of quantum chromodynamics (QCD), binds quarks into hadrons and generates the majority of hadron mass. 
Important information concerning the quark dynamics inside hadrons is obtained from their intrinsic electromagnetic structure and described by electromagnetic form factors~\cite{EMFFsProton}. 
While space-like form factors for the proton and neutron are accessible through elastic electron scattering, the most viable option for unstable hadrons is the time-like form factors.
Recently, precise measurements of pair production of protons~\cite{BaBarppbar,BESIIIppbar2015,BESIIIppbar2019,BESIIIppbar2020,BESIIIppbar2021}, neutrons~\cite{BESIIInnbar,SNDnnbar,BESIIInnbarR}, and strange hyperons~\cite{BaBarLLbar,BESIIILLbar2018,BESIIILLbar2020,BESIIILLbar2021,BESIIILLbar2022,BESIIILLbar2023} in the annihilation of electron and positron has brought renewed insights into the electromagnetic structure of baryons~\cite{Huang:2021xte}. 
In these measurements, the non-zero cross sections near kinematic threshold, followed by a wide-range plateau, have triggered various theoretical interpretations~\cite{ppbarfsi,ppbarCfactor1,ppbarCfactor2}. 
Moreover, a striking oscillation feature has been observed in the energy-dependence of the effective form factor of the proton~\cite{ppbarOsc2015}. 
This feature has been confirmed for the neutron with the same oscillation frequency by BESIII~\cite{BESIIInnbar}, while a recent SND measurement near threshold suggests a much lower frequency~\cite{SNDnnbar}. 
A similar oscillatory behavior has also been extracted from the proton $\gEgM$ distribution~\cite{ppbarOsc2021}. 
The source of the oscillations is not yet established but has been discussed extensively~\cite{NNbarRev2021,NNbarRev2022,NNbarRes2022,NNbarThreshold2022}. 

As the charm analogue of proton, the $\lam$ hadron can shed new light on  baryon structure. 
The pair production process of $\eelplm$ was firstly studied by Belle via the initial-state radiation (ISR) technique~\cite{Bellelplm}. 
A resonant structure around the center-of-mass (c.m.) energy ($\sqs$) of $4.63~\gev$, denoted as the $Y(4630)$, was discerned in the cross-section line shape. 
This charmonium-like state is regarded as an exotic-hadron candidate, such as a charmed baryonium~\cite{CharmedBaryonium,Lee:2011rka}, a meson-meson molecular state~\cite{Guo:2010tk} or a tetraquark state~\cite{Maiani:2014aja,Liu:2016sip}. 
However, the BESIII measurement of $\eelplm$ from the threshold to 4.6~$\gev$~\cite{BESIIIlplm} implies a different energy-dependence trend of the cross section. 
This significantly affects the parametrization of the $Y(4630)$~\cite{BesBellDiff1,BesBellDiff2,BesBellDiff3,BesBellDiff4}. 
To understand the interplay between the $\lam$ pair production and the charmonium-like resonance, a high-precision measurement of the cross section around 4.63~$\gev$ is required~\cite{Interplay1,Interplay2,Reviewlplm}. 
A yet deeper understanding of these dynamics can be gained by investigating the energy-dependence of $\gEgM$~\cite{gEgMTheo1,gEgMTheo2}.
This has only been measured at two points below 4.6~$\gev$ by BESIII~\cite{BESIIIlplm}. 
Therefore, a thorough study of the effective form factor and $\gEgM$ ratio of the $\lam$ baryon will greatly contribute to the interpretation of the oscillation features and  baryon structure.

In this Letter, the Born cross sections ($\sigma$) of the process $\eelplm$ and the effective form factor of $\lam$ are determined at twelve c.m. energies from $\ENERGYAT$ to $\ENERGYLT~\gev$~\cite{lumixyz}.
In addition, the Born polar angle ($\theta$) distribution of $\lam$ is analyzed at each energy point to determine $\gEgM$ and from that, $\GM$. 
The polar angle of the $\lam$ baryon, which is produced by a virtual photon ($\gamvir$), assuming one-photon exchange in electron-positron annihilation, is defined as the angle between the momenta of $\lam$ and positron in the rest frame of $\gamvir$. 
In this work,  the decay mode $\lam\to\modeI$ and its charge conjugate (referred to as the signal mode hereafter) are employed to reconstruct the $\lam$ and $\lambar$ signals, respectively.   This mode benefits from a relatively large branching fraction (BF)~\cite{PDG2022}.
As a result, two individual measurements of the cross section and the polar-angle distribution are obtained at each c.m. energy, which are then averaged to yield the final result. 
To reduce the systematic uncertainty related to the BF of the signal mode and detection efficiency, a double-tag (DT) approach, where both the $\lam$ and $\lambar$ baryons are reconstructed in each event, is used in addition to the single-tag (ST) method~\cite{BESIIIlplm}.

The data were collected with the BESIII detector~\cite{BESIII} operating at BEPCII. Monte Carlo (MC) packages based on the {\sc geant}4 software~\cite{geant4} are used to produce simulated events, where the interaction between secondary particles and the detector material is included. 
To estimate the detection efficiency, the {\sc kkmc}~\cite{KKMC} program is used to generate the ST and DT signal MC events, where the ISR~\cite{KKMCISR} and beam-energy spread~\cite{EneSpd} effects are simulated. 
In the signal MC samples, the tagged $\lampm$ is set to decay via the signal mode which is modeled by the dedicated partial-wave analysis, while the untagged one decays inclusively according to the BFs listed in Particle Data Group (PDG)~\cite{PDG2022,SpeBRMod}. 
In addition, the c.m. energy-dependent Born cross section and polar angle distribution of $\lam$, which are measured and parameterized in this work, are implemented in {\sc kkmc} iteratively. 
To study the background, inclusive MC samples, including the $\lplm$, QED-related, and hadronic~\cite{Hybrid} (with the $\lplm$ events excluded) events, are produced. The subsequent decays of all the intermediate states in MC samples are simulated by {\sc evtgen}~\cite{EVTGEN}.

The $\lampm$ candidates are formed with the charged tracks selected with the same criteria as those used in Ref.~\cite{BESIIIBRs}. 
The energy difference $\dele=E-\ebm$ and beam-constrained mass $\mbc=\sqrt{\ebm/c^{4}-p^{2}/c^{2}}$ are utilized to determine the number of the ST signal events, where $E$ and $p$ are the energy and momentum of the $\lampm$ candidates, respectively. 
The same approach, as described in Ref.~\cite{BESIIIlplm}, is applied here on the ST signal candidates at all the c.m. energies, except for an asymmetric requirement window of $(-34, 20)~\mev$ for $\dele$. 
Studies based on the signal and inclusive MC samples demonstrate that the simulation reproduces experimental data well and the background in the $\mbc$ distribution can be described by an ARGUS function~\cite{ARGUSf}. 
The ST yield ($\nst$) and detection efficiency ($\effst$) are determined by applying unbinned maximum likelihood fits on the $\mbc$ distributions of data and the ST signal MC samples, respectively. 
The fit result at $\sqs=\ENERGYET~\gev$ is shown in Fig.~\ref{lamcSTsignal}, and those at other c.m. energies can be found in the Supplemental Material~\cite{SupMat}.

\begin{figure}[!htbp]
\setlength{\abovecaptionskip}{-0.0cm}
\setlength{\belowcaptionskip}{-0.4cm}
\begin{center}
\begin{overpic}[width=2.91in]{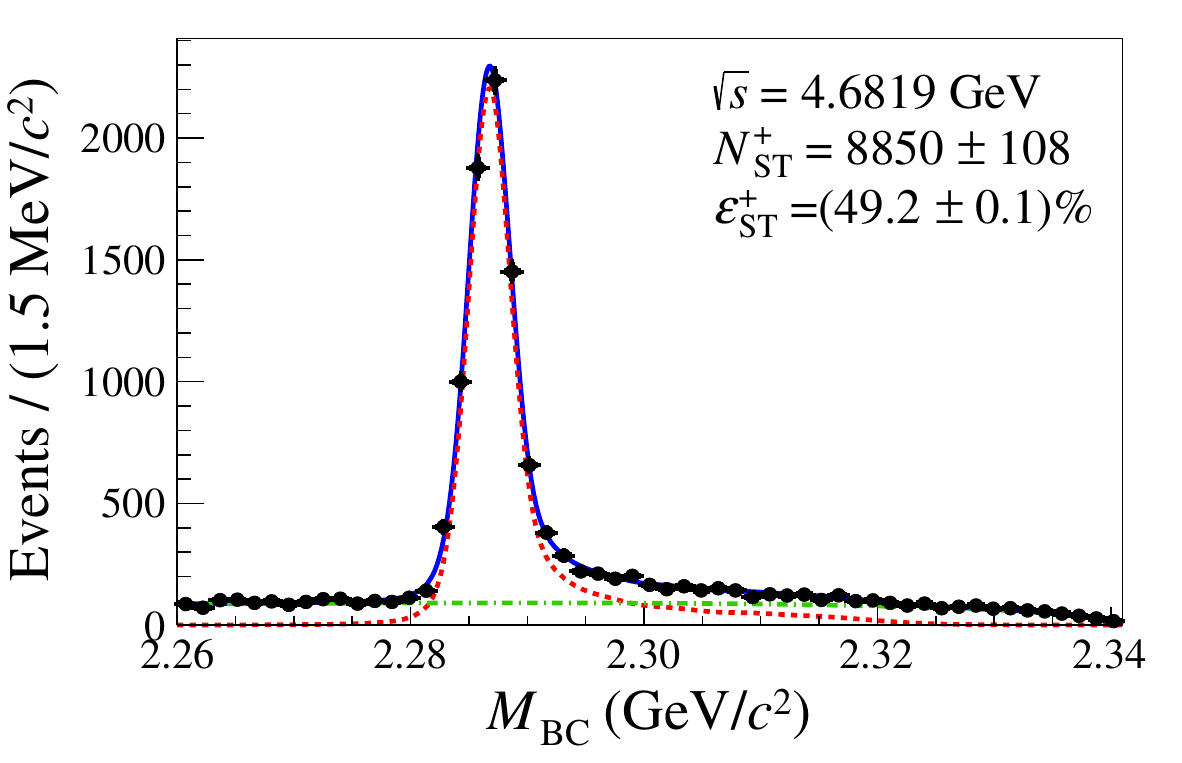}
\end{overpic}
\caption{Fit to the $\mbc$ distribution of the $\lam$ candidates at $\sqs=\ENERGYET~\gev$, where the black dots denote data and the blue solid curve is the sum of fit functions. The signal and background functions are illustrated by the red dashed and green dash-dotted curves, respectively. Here, $\nstp$ and $\effstp$ are the ST yield and detection efficiency, where the uncertainties are statistical.}
\label{lamcSTsignal}
\end{center}
\end{figure}


The Born cross section of the $\eelplm$ process is calculated using
\begin{equation}
\cspm = \frac{\nstpm}{\effstpm\fisr\fvp\lint\brfpm},
\label{csformula}
\end{equation}
where the indexes ``$\pm$'' denote the positive and negative ST modes. The ISR correction factor $\fisr$ is derived from the QED theory~\cite{KKMCISR} and 
calculated with {\sc kkmc} in an iterative process where the Born cross-section line shape is used as input and updated in each iteration.
The vacuum polarization (VP) correction factor $\fvp$ is between 1.054 and 1.056 in this energy region~\cite{VPfactor}. 
The integrated luminosity $\lint$ is measured with Bhabha events at each energy point~\cite{lumixyz}. 
The BFs of the signal mode and its charge conjugate, denoted as $\brfpm$, are known to a precision of 5.1\%~\cite{PDG2022} which would give rise to a significant uncertainty in the cross section. 
To avoid this problem, a DT analysis is carried out using the data sets with $\lint$ greater than $350~\textmd{pb}^{-1}$.  
The total number of the DT events ($\ndt$) is proportional to $\brfpm$ as~\cite{SupMat}
\begin{equation}
\ndt = \brfpm\sum_{n=1}^{9}\bigg(\frac{\nst^{\mp,n}\effdt^{n}}{\effst^{\mp,n}}\bigg),
\label{brfpmform}
\end{equation}
where nine data sets, including that at $\sqs=\ENERGYPT~\gev$, are analyzed and $\effdt$ is the DT detection efficiency. The superscript ``$n$'' indicates the $n$-th c.m. energy used in the DT analysis. Because of the statistical fluctuations of $\nst^{\mp,n}$, $\brfp$ and $\brfm$ are considered to be separate observables. 
Combining Eq.~(\ref{csformula}) and Eq.~(\ref{brfpmform}), the individual cross section is recast as
\begin{equation}
\cspm = \frac{\nstpm}{\effstpm\fisr\fvp\lint\ndt}\sum_{n=1}^{9}\bigg(\frac{\nst^{\mp,n}\effdt^{n}}{\effst^{\mp,n}}\bigg),
\label{csnewformula}
\end{equation}
where the ratio $\effdt^{n}/(\effst^{\mp,n}\effstpm)$ cancels most of the systematic effects caused by the efficiency differences between data and MC simulation in tracking and particle identification (PID). 

Candidate DT events are selected with the following criteria: 
(i) at least one proton, $K^{-}$ and $\pi^{+}$ meson, as well as their charge conjugates, are selected and identified in each event via the same tracking and PID requirements as applied in the ST analysis; 
(ii) the $\lam$ and $\lambar$ candidates are reconstructed with the decays $\lam\to\modeI$ and $\lambar\to\amodeI$, respectively. 
If there is more than one $\lam$($\lambar$) candidate in an event, the one with the smallest $|\dele_{+}|$($|\dele_{-}|$) is retained; 
(iii) the same acceptance window as used in the ST analysis is implemented on $|\delepm|$ at all the nine c.m. energies to suppress the background from non-signal $\lampm$ decays. 
These selection criteria are also applied on the DT signal MC sample to evaluate the corresponding $\effdt$~\cite{SupMat}. 

After the above selection, the two-dimensional (2D) distribution, i.e., $(\mbcp,~\mbcm)$, is obtained at each c.m. energy. 
Studies with MC samples demonstrate that the simulated DT events reproduce data well and 
the contamination from $\lplm$ events with non-signal decays is negligible. 
The dominant background comes from the inclusive hadronic events, which accumulates in 
the vicinity of $\mbcp=\mbcm$. 
To determine $\ndt$, a 2D simultaneous unbinned likelihood fit is applied on the ($\mbcp,~\mbcm$) distributions of the nine data sets, where the BF $\brfpm$ is the shared parameter at all the c.m. energies. 
In each ($\mbcp,~\mbcm$) distribution, the signal is described by the corresponding DT signal MC shape convoluted with a 2D Gaussian function. 
The background is described by the product of an ARGUS function and a Gaussian function with the arguments $(\mbcp+\mbcm)/2$ and $(\mbcp-\mbcm)$, respectively. 
In the background functions, the truncation parameter of the ARGUS function is fixed to be the corresponding $\ebm$, while the other parameters are free and shared by different c.m. energies. 
There are two simultaneous fits sharing $\brfp$ and $\brfm$, respectively. The one-dimensional projections of the two 2D fits at different c.m. energies are shown in Ref.~\cite{SupMat}. 
From these fits, the BFs are determined to be $\brfp=\brfpval$ and $\brfm=\brfmval$, which give an average of $\brf=\brfval$. 
Here the uncertainties are statistical, and $\brfp$ and $\brfm$ are fully correlated. 
The DT approach is validated by an MC study using the combined inclusive $\lplm$ and hadronic MC samples, which reproduce the signal and background processes, respectively. 
These inclusive MC samples are ten times the size of the data samples. 
Moreover, the $\brfpm$ are evaluated individually at each c.m. energy by the similar 2D fit. 
No energy-dependency is observed in the nine groups of $\brfpm$ and their weighted averages are consistent with the values obtained with the two simultaneous 2D fits.

From the DT analysis, the total DT yield is determined to be $\ndt=\ndtval$. Accordingly, the individual cross sections at each c.m. energy are determined with Eq.~(\ref{csnewformula}), which are given in the Supplemental Material~\cite{SupMat}.

The systematic uncertainties on the cross-section measurement come from reconstruction-related and general sources. 
The former is mainly due to the size of the signal MC samples, the MC modeling of the $\lampm$ production and decay, the tracking and PID efficiencies of final-state particles, and the DT analysis. 
The uncertainty of $\effst$ arising from the limited MC sample size, which varies from 0.1\% to 0.2\% for different c.m. energies, is taken as the systematic uncertainty. 
At higher energies, $\lampm$ is usually produced with higher momentum, therefore the rest frame of its decay products is highly boosted. 
Since the detection efficiency of $\lampm$ with small scattering angle decreases due to the limited acceptance at the edge of the detector, the uncertainty of the polar-angle distribution input into {\sc kkmc} propagates into $\effst$ and $\effdt$, and thereby the cross section. 
These systematic uncertainties are estimated to be less than 0.6\%. 
The MC modeling of the signal mode is validated by extensive comparisons between data and MC simulation, and is considered to have a negligible contribution to the systematic uncertainty.
Although the DT procedure is intrinsically robust against systematic bias, there is still a residual uncertainty associated with the tracking and PID efficiencies. 
Studies based on the control samples of $J/\psi\to\ppbar\pipi$ and $J/\psi\to\ks K^{\pm}\pi^{\mp}$ decays, are used to correct $\effst$ and $\effdt$, and re-evaluate the cross sections.
The observed relative differences in $\csp$ and $\csm$, which are less than 0.4\% and 0.1\%, respectively, are taken as the systematic uncertainties. 

The systematic uncertainty associated with the DT analysis has three components: (i) the statistical uncertainty of $\ndt$ which is determined to be 3.2\% from the 2D simultaneous fit; (ii) the description of the background component in the simultaneous 2D fit, for which two alternative background functions are tested; (iii) the uncertainties of $\nst^{\mp,n}$, $\effst^{\mp,n}$, and $\effdt^{n}$ appearing in Eq.~(\ref{csnewformula}). 
The total uncertainty on the cross section from these sources is 3.3\%, which is less than the 5.1\% uncertainties on $\brfpm$ according to PDG~\cite{PDG2022}. This is the reason we implement the DT approach in this analysis.

The systematic uncertainties on the cross section associated with the $\dele$ and $\mbc$ requirements are negligible since the signal MC sample reproduces the data well. Moreover, the fit model of $\mbc$ in the ST analysis does not introduce any significant systematic uncertainty.

The general sources that contribute to the systematic uncertainties on the cross section arise from the evaluations of $\fisr$, $\fvp$, and $\lint$. 
By using different calculation algorithms, inputting alternative cross section line-shapes in the {\sc kkmc} generator, and considering the uncertainties of the c.m. energy~\cite{lumixyz} and energy spread~\cite{EneSpd}, the total uncertainty of $\fisr$ is estimated to be 2.3\% at $\sqs=\ENERGYGT~\gev$ and lower than 1.0\% at all other energy points. 
The uncertainty on $\fvp$ is assigned to be 0.5\%~\cite{VPfactor} and that of $\lint$ is about 0.5\%~\cite{lumixyz} at all the c.m. energies. 

All the systematic uncertainties of $\csp$ and $\csm$ are correlated at the same c.m. energy, except for those arising from the statistical uncertainties of $\nst$, $\effst$, and $\effdt$.  
Furthermore, the systematic uncertainties from the DT analysis, $\fvp$, and $\lint$, obtained at different energy points, are correlated.  
Details of these systematic uncertainties are tabulated in Ref.~\cite{SupMat}.

\begin{figure}[!htbp]
\setlength{\abovecaptionskip}{-0.0cm}
\setlength{\belowcaptionskip}{-0.4cm}
\begin{center}
\begin{overpic}[width=3.3in]{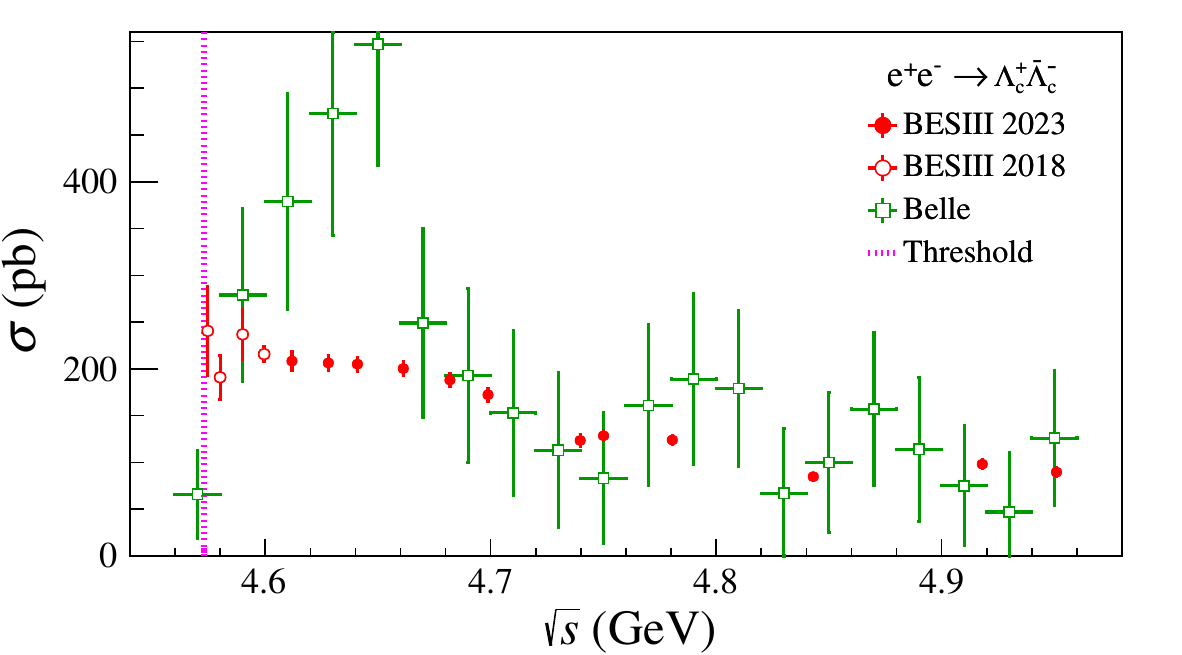}
\end{overpic}
\caption{Comparison of the cross sections of the $\eelplm$ process, where the red dots denote the results of this study and the green open squares indicate those of Belle~\cite{Bellelplm}. The results of the previous BESIII measurement~\cite{BESIIIlplm} are also updated and shown as red open dots.}
\label{lamccrosecplot}
\end{center}
\end{figure}

At each c.m. energy, the average cross section is determined with the method described in Ref.~\cite{BESIIIlplm,AverageData}. The results are presented in Table~\ref{lamctab}. 
Figure~\ref{lamccrosecplot} illustrates the comparison of the $\eelplm$ cross sections measured in this study, and by Belle~\cite{Bellelplm}. Also shown are the results of the previous BESIII measurements~\cite{BESIIIlplm}, which have been re-evaluated using the updated variables required in Eq.~(\ref{csnewformula})~\cite{SupMat}. In our data, the near threshold cross-section plateau is confirmed up to 4.66~$\gev$ and no resonance structure is observed around 4.63~$\gev$.

The effective $\lam$ form factor is calculated from the average cross section $\sigma$ as
\begin{equation}
\Geff=\sqrt{\frac{\sigma}{\frac{\sigma_{0}}{3}\big(1+\frac{\kappa}{2}\big)}},
\label{geffexpress}
\end{equation}
where $\sigma_{0} = 4\pi\alpha^{2}\beta C/s$, $C$ is the Coulomb factor~\cite{BESIIIlplm}, $\beta=\sqrt{1-\kappa}$, $\kappa=4m^{2}c^{4}/s$, and $m$ is the known mass of the $\lam$ baryon~\cite{PDG2022}. 
Table~\ref{lamctab} lists the calculated $\Geff$ above 4.6~$\gev$ while those near threshold are given in Ref.~\cite{SupMat}. 
The three-pole model~\cite{ppbarOsc2021} is used to fit the $\Geff$ distribution, 
where an oscillatory behavior is expected in the residuals between data and the fitted model. 
However, neither the model nor its variants~\cite{ppbarOsc2015} can describe the $\Geff$ distribution. 
In addition, there is no discernible oscillation feature in the residual distribution~\cite{SupMat}. 


To precisely determine the $\gEgM$ value for $\lam$ production at a given c.m. energy, the Born polar-angle distribution of $\lam$ production is studied~\cite{diffcrosecs} using the ST signal sample. 
There is a sizable fraction of $\lplm$ ISR-return events in the ST signal sample, for which the polar angle of $\lam$ is not accessible.  
However, for pure Born events, the polar angle coincides with the scattering angle. Therefore, the Born polar-angle distribution can be obtained by applying a $\costh$-dependent correction, accounting for ISR effects, on the produced scattering angle distribution. Based on the ST signal MC sample, where the ISR events can be distinguished, the correction is obtained by dividing the normalized generated scattering-angle distribution of all the ST signal events by that of the ST sample with the ISR events excluded. 
The ISR correction is further parameterized by an empirical function to achieve a smooth $\costh$-dependent correction. 

Benefiting from the large ST yields~\cite{SupMat} at each c.m. energy, the ST sample is divided into 20 $\costh$ bins~\cite{Scattering}. In each bin, the ST yield is obtained via a fit to the corresponding $\mbc$ spectrum. 
The one-dimensional bin-by-bin efficiency and ISR corrections are successively applied on these $\costh$-dependent ST yields to obtain the individual Born polar-angle distributions. 
Then the average Born polar-angle distribution of $\lam$ is fitted with the function $f(\costh)=N_{0}(1+\alplam\cossqth)$, where $N_{0}$ is proportional to the average Born cross section. 
The shape parameter $\alplam=(1-\kappa R^{2})/(1+\kappa R^{2})$, where $R=\gEgM$~\cite{diffcrosecs}. 
The fit results are shown in Ref.~\cite{SupMat}. 
The obtained $\alplam$ and $\gEgM$ are listed in Table~\ref{lamctab}. 
The modulus of the magnetic form factor $\GM^{2}$ is evaluated using
\begin{equation}
\GM^{2}=\frac{2\nbin}{\sigma_{0}\fisr\fvp\lint\brf}N_{0}(1+\alplam),
\label{gmexpress}
\end{equation}
where $\nbin=20$ and $\brf$ is the average BF given previously. 
The reliability of the method is validated by studying a ST signal MC sample which is of a size 100 times larger than that of data.

The systematic uncertainty of $\alplam$, which propagates to that of $\gEgM$ and $\GM$, is addressed source-by-source. 
Using the tracking and PID efficiencies obtained from the aforementioned control samples, $\effst$ in each $\costh$ bin is corrected and  $\alplam$ is re-evaluated.
The resulting differences, which are typically less than 1.2\%~\cite{AngDisSysNote}, are regarded as the systematic uncertainties. 
The uncertainties of $\alplam$ arising from the signal migration between different $\costh$ bins are found to be smaller than 4.7\%. 
Since the size of the ST signal MC sample is limited, there is uncertainty in the parameters of the empirical ISR correction function. 
These parameters are changed by the size of the corresponding uncertainty to estimate the systematic uncertainty of $\alplam$, for which 5.5\% is obtained at most. 
The systematic uncertainties due to the $\dele$ requirement, the $\mbc$ fit, the MC modeling of the signal decay, the bin size, and fit range of $\costh$ are negligible. 

\begin{figure}[!htbp]
\setlength{\abovecaptionskip}{-0.0cm}
\setlength{\belowcaptionskip}{-0.4cm}
\begin{center}
\begin{overpic}[width=3.3in]{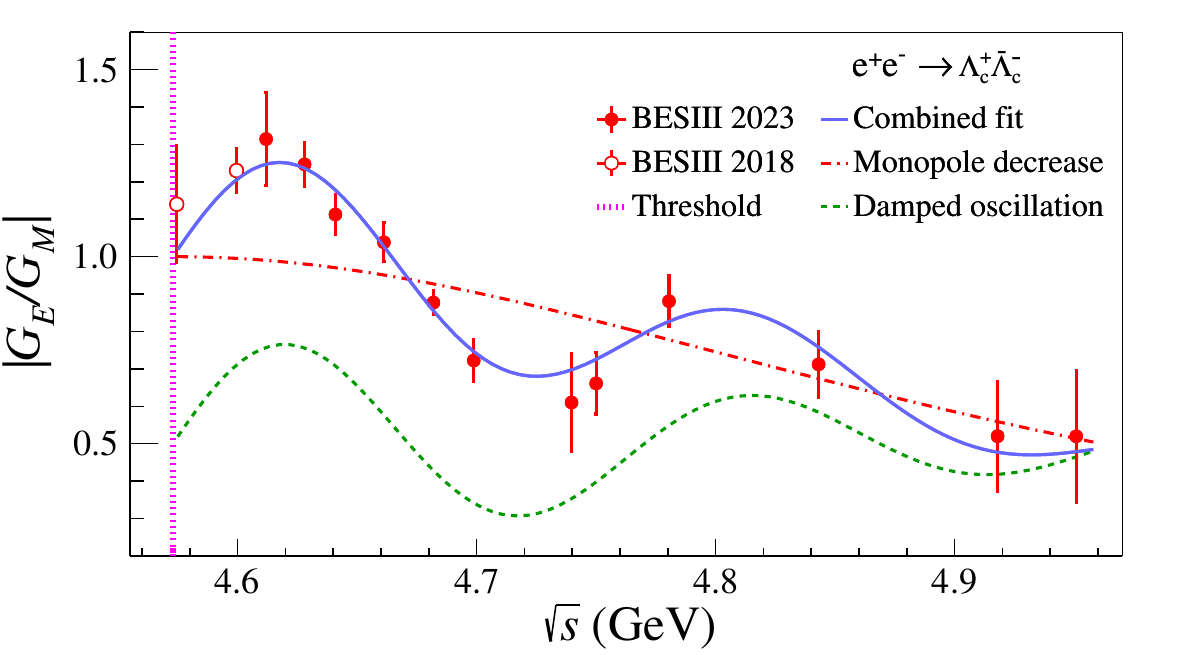}
\end{overpic}
\caption{Measured $\gEgM$ of the charmed baryon in the $\eelplm$ process, where the red dots denote the results of this study and the red open circles indicate those of Ref.~\cite{BESIIIlplm}. The blue solid curve represents a fit consisting of a damped oscillation (green dashed line after a shift by 0.5 in $\gEgM$) on top of the monopole decrease (red dash-dotted curve).}
\label{lamcgegmplot}
\end{center}
\end{figure}

Table~\ref{lamctab} lists the measured $\alplam$, $\gEgM$, and $\GM$, where the systematic uncertainty of $\GM$ includes the contributions from the uncertainties of the variables in the denominator of Eq.~(\ref{gmexpress}). 
Figure~\ref{lamcgegmplot} shows the resulting $\gEgM$ obtained in this work and the previous BESIII measurement~\cite{BESIIIlplm}. 
The figure also illustrates a fit using a function combining the monopole decrease with a damped oscillation~\cite{ppbarOsc2021}:
\begin{equation}
\gEgM(s)=\frac{1}{1+\omega^{2}/r_{0}}\big[1+r_{1}e^{-r_{2}\omega}\sin(r_{3}\omega)\big],
\label{fitcrosecs}
\end{equation}
where $\omega=\sqs-2m$ and $r_i$ with $i=0,1,2,3$ are free parameters. 
The oscillation frequency is determined to be $r_3=(32\pm1)~\gev^{-1}$, which is about 3.5 times greater than that measured for the proton~\cite{ppbarOsc2021}.

\linespread{1.0}
\begin{table*}[!htbp]
\setlength{\abovecaptionskip}{3pt}
\setlength{\belowcaptionskip}{-5pt}
\caption{Summary of the measured average production cross section, effective form factor, polar angle distribution parameter, electromagnetic form factor ratio, and magnetic form factor of the charmed baryon in the $\eelplm$ process at each energy point, where the uncertainties are statistical and systematic, respectively.}
\label{lamctab}
\begin{center}
\setlength{\tabcolsep}{2.6mm}{
\begin{tabular}{ c  r  r  c  r  c  c }
\hline
\hline
\multicolumn{1}{  c  }{\multirow{1}{*}{$\sqs$} } &
\multicolumn{1}{  c  }{\multirow{1}{*}{$\lint$} } &
\multicolumn{1}{  c  }{\multirow{2}{*}{$\sigma$~(pb)} } &
\multicolumn{1}{  c  }{\multirow{2}{*}{$\Geff$~($10^{-2}$)} } &
\multicolumn{1}{  c  }{\multirow{2}{*}{$\alplam$} } &
\multicolumn{1}{  c  }{\multirow{2}{*}{$\gEgM$} } &
\multicolumn{1}{  c  }{\multirow{2}{*}{$\GM$~($10^{-2}$)} } \\
\multicolumn{1}{  c  }{\multirow{1}{*}{($\gev$)} } &
\multicolumn{1}{  c  }{\multirow{1}{*}{(pb$^{-1}$)} } &
\multicolumn{1}{  c  }{\multirow{1}{*}{} } &
\multicolumn{1}{  c  }{\multirow{1}{*}{} } &
\multicolumn{1}{  c  }{\multirow{1}{*}{} } &
\multicolumn{1}{  c  }{\multirow{1}{*}{} } &
\multicolumn{1}{  c  }{\multirow{1}{*}{} } \\ \cline{1-7}
\hline
\ENERGYAT  &  \lintA~  &  $\PkpiBornCSCA$  &  $\PkpiBornGefA$  &  $\pkpiFinalpHaA$   & $\pkpiFinRatioA$  & $\pkpiFinGmmodA$ \\ 
\ENERGYBT  &  \lintB~  &  $\PkpiBornCSCB$  &  $\PkpiBornGefB$  &  $\pkpiFinalpHaB$   & $\pkpiFinRatioB$  & $\pkpiFinGmmodB$ \\ 
\ENERGYCT  &  \lintC~  &  $\PkpiBornCSCC$  &  $\PkpiBornGefC$  &  $\pkpiFinalpHaC$   & $\pkpiFinRatioC$  & $\pkpiFinGmmodC$ \\ 
\ENERGYDT  &  \lintD~  &  $\PkpiBornCSCD$  &  $\PkpiBornGefD$  &  $\pkpiFinalpHaD$   & $\pkpiFinRatioD$  & $\pkpiFinGmmodD$ \\ 
\ENERGYET  &  \lintE~  &  $\PkpiBornCSCE$  &  $\PkpiBornGefE$  &  $\pkpiFinalpHaE$   & $\pkpiFinRatioE$  & $\pkpiFinGmmodE$ \\ 
\ENERGYFT  &  \lintF~  &  $\PkpiBornCSCF$  &  $\PkpiBornGefF$  &  $\pkpiFinalpHaF$   & $\pkpiFinRatioF$  & $\pkpiFinGmmodF$ \\ 
\ENERGYGT  &  \lintG~  &  $\PkpiBornCSCG$  &  $\PkpiBornGefG$  &  $\pkpiFinalpHaG$   & $\pkpiFinRatioG$  & $\pkpiFinGmmodG$ \\ 
\ENERGYHT  &  \lintH~  &  $\PkpiBornCSCH$  &  $\PkpiBornGefH$  &  $\pkpiFinalpHaH$   & $\pkpiFinRatioH$  & $\pkpiFinGmmodH$ \\ 
\ENERGYIT  &  \lintI~  &  $\PkpiBornCSCI$  &  $\PkpiBornGefI$  &  $\pkpiFinalpHaI$   & $\pkpiFinRatioI$  & $\pkpiFinGmmodI$ \\ 
\ENERGYJT  &  \lintJ~  &  $\PkpiBornCSCJ$  &  $\PkpiBornGefJ$  &  $\pkpiFinalpHaJ$   & $\pkpiFinRatioJ$  & $\pkpiFinGmmodJ$ \\ 
\ENERGYKT  &  \lintK~  &  $\PkpiBornCSCK$  &  $\PkpiBornGefK$  &  $\pkpiFinalpHaK$   & $\pkpiFinRatioK$  & $\pkpiFinGmmodK$ \\ 
\ENERGYLT  &  \lintL~  &  $\PkpiBornCSCL$  &  $\PkpiBornGefL$  &  $\pkpiFinalpHaL$   & $\pkpiFinRatioL$  & $\pkpiFinGmmodL$ \\ 
\hline
\hline
\end{tabular}}
\end{center}
\end{table*}

In summary, the Born cross sections and polar angle distributions of the process $\eelplm$ are studied at twelve c.m. energies from $\ENERGYAT$ to $\ENERGYLT~\gev$. 
Benefiting from the large data samples, which enable ST and DT approaches via the decay $\lam\to\modeI$, the cross sections and effective form factors of $\lam$ are determined with an unprecedented precision. 
From the threshold up to $4.66~\gev$, our measured cross sections indicate no enhancement around the $Y(4630)$ resonance, which is different from Belle~\cite{Bellelplm}. 
In contrast to the case for the proton and neutron, no oscillatory behavior is discerned in the effective form-factor spectrum of $\lam$. 
However, the energy-dependence of $\gEgM$ reveals an oscillation feature with a 
significantly higher frequency than that of the proton, which 
may imply a non-trivial structure of the lightest charmed baryon.

The BESIII Collaboration thanks the staff of BEPCII, the IHEP computing center and the supercomputing center of USTC for their strong support. This work is supported in part by National Key R\&D Program of China under Contracts Nos. 2020YFA0406400, 2020YFA0406300; National Natural Science Foundation of China (NSFC) under Contracts Nos. 11635010, 11735014, 11835012, 11935015, 11935016, 11935018, 11961141012, 12022510, 12025502, 12035009, 12035013, 12061131003, 12192260, 12192261, 12192262, 12192263, 12192264, 12192265, 12221005, 12225509, 12235017; China Postdoctoral Science Foundation under Contracts No. 2019M662152, No. 2020T130636; the Fundamental Research Funds for the Central Universities, University of Science and Technology of China under Contract No. WK2030000053; the Chinese Academy of Sciences (CAS) Large-Scale Scientific Facility Program; the CAS Center for Excellence in Particle Physics (CCEPP); Joint Large-Scale Scientific Facility Funds of the NSFC and CAS under Contracts No. U1732263, No. U1832207, No. U1832103, No. U2032105, No. U2032111; CAS Key Research Program of Frontier Sciences under Contracts Nos. QYZDJ-SSW-SLH003, QYZDJ-SSW-SLH040; 100 Talents Program of CAS; The Institute of Nuclear and Particle Physics (INPAC) and Shanghai Key Laboratory for Particle Physics and Cosmology; ERC under Contract No. 758462; European Union's Horizon 2020 research and innovation programme under Marie Sklodowska-Curie grant agreement under Contract No. 894790; German Research Foundation DFG under Contracts Nos. 443159800, 455635585, Collaborative Research Center CRC 1044, FOR5327, GRK 2149; Istituto Nazionale di Fisica Nucleare, Italy; Ministry of Development of Turkey under Contract No. DPT2006K-120470; National Research Foundation of Korea under Contract No. NRF-2022R1A2C1092335; National Science and Technology fund of Mongolia; National Science Research and Innovation Fund (NSRF) via the Program Management Unit for Human Resources \& Institutional Development, Research and Innovation of Thailand under Contract No. B16F640076; Polish National Science Centre under Contract No. 2019/35/O/ST2/02907; The Swedish Research Council; U. S. Department of Energy under Contract No. DE-FG02-05ER41374.




\begin{thebibliography}{99}
\bibitem{SpeBRMod} To be accurate, for the inclusive decay of the charmed baryon, $\brf$ of the signal mode is set to be the one obtained in this work and all the available partial wave analysis results are used to model the corresponding decay channels.
\bibitem{Scattering} Here, the binning is carried out in terms of the scattering angle.
\bibitem{AngDisSysNote} The relative systematic uncertainties of $\alplam$ at $\sqs=\ENERGYCT$ and $\ENERGYDT~\gev$ are significantly greater than those at the other c.m. energies due to the corresponding vanishing $\alplam$. While the total absolute systematic uncertainties of these two $\alplam$ are less than 0.01. Therefore, they are not counted in the maximum relative uncertainties given in this Letter.
\end{thebibliography}
\end{document}